# PICASO 4.0: Clouds and Photochemistry in Climate Models of Brown Dwarfs and Exoplanets

James Mang[1,13], Natasha E. Batalha[2], Caroline V. Morley[1], Nicholas F. Wogan[2,3], Sagnick Mukherjee[4,14], Channon Visscher[5,6], Mark S. Marley[7], Jonathan J. Fortney[8], Katy L. Chubb[9,10], Peter Gao[11], and Isaac Malsky[12]

[1] Department of Astronomy, University of Texas at Austin, Austin, TX 78712, USA; j_mang@utexas.edu
[2] NASA Ames Research Center, Moffett Field, CA 94035, USA
[3] SETI Institute, Mountain View, CA 94043, USA
[4] School of Earth and Space Exploration, Arizona State University, Tempe, AZ, USA
[5] Chemistry & Planetary Sciences, Dordt University, Sioux Center, IA 51250, USA
[6] Center for Extrasolar Planetary Systems, Space Science Institute, Boulder, CO 80301, USA
[7] Department of Lunar and Planetary Sciences, University of Arizona, Tucson, AZ 85721, USA
[8] Department of Astronomy & Astrophysics, University of California Santa Cruz, Santa Cruz, CA 95064, USA
[9] Centre for Exoplanet Science, University of St Andrews, North Haugh, Saint Andrews, KY16 9SS, UK
[10] University of Bristol, School of Physics, Tyndall Avenue, Bristol, BS8 1TL, UK
[11] Earth & Planets Laboratory, Carnegie Institution for Science, 5241 Broad Branch Road Northwest, Washington, DC 20015, USA
[12] Jet Propulsion Laboratory, California Institute of Technology, Pasadena, CA 91109, USA

## Abstract

We present a major update to the open-source atmospheric modeling package PICASO, designed for simulating the thermal structure and spectra of hydrogen-rich atmospheres of brown dwarfs and exoplanets. This release, PICASO 4.0, expands upon the existing radiative-convective equilibrium model framework by incorporating several new capabilities. Key additions include the integration of Virga for self-consistent cloud modeling, new flexible treatments for rainout and cold trapping of volatile species, and support for photochemistry. We also introduce a parameterized energy injection scheme to simulate additional external or internal heating processes. These features are motivated by lessons from recent JWST observations that reveal the prevalence of nonequilibrium chemistry and clouds. We benchmark the new functionalities against previously published results in the literature, including the Sonora Diamondback grid, energy injected atmospheres, patchy cloud models, and other photochemical models of WASP-39b. PICASO continues to be actively developed as an open-source package aimed at enabling reproducible, community-driven atmospheric modeling of all substellar objects.

## 1. Introduction

With JWST now delivering high-precision observations across a wide range of substellar objects, it has become clear that their atmospheres are shaped by complex and often nonequilibrium chemistry. Brown dwarfs and giant planets show evidence for strong vertical mixing (A. J. Skemer et al. 2014; B. E. Miles et al. 2020; E. C. Matthews et al. 2024; D. K. Sing et al. 2024; L. Welbanks et al. 2024; S. Barat et al. 2025; D. C. Bardalez Gagliuffi et al. 2025) and signatures of clouds (J. K. Faherty et al. 2014; C. V. Morley et al. 2018; B. Benneke et al. 2019; D. Grant et al. 2023; B. E. Miles et al. 2023; J. Inglis et al. 2024; G. Fu et al. 2025; K. K. W. Hoch et al. 2025; S. Mukherjee et al. 2025c) which cannot be captured with clear atmospheric models in chemical equilibrium. One-dimensional radiative-convective equilibrium (RCE) models remain a core tool for providing a physically consistent model of the thermal structures, emergent spectra, and evolution of H/He-dominated atmospheres. Increasing the complexity and flexibility of forward modeling frameworks is essential to improve our characterization of the growing population of cold giant planets detected with JWST (R. Bowens-Rubin et al. 2025) and prepare for reflected-light studies with the Roman/CGI (R. E. Lupu et al. 2016) and direct imaging with the upcoming ELTs.

Several open-source RCE codes have been developed, each with unique features and applications. ATMO (P. Tremblin et al. 2015; M. W. Phillips et al. 2020) includes flexible, nonadiabatic thermal structures and supports both chemical equilibrium and disequilibrium. petitCODE (P. Mollière et al. 2015, 2017) provides climate models in chemical equilibrium, computes transmission spectra, and can include parameterized clouds. HELIOS (M. Malik et al. 2017) is a GPU-accelerated RCE model that assumes chemical equilibrium. MSG (B. Campos Estrada et al. 2025) incorporates a microphysical cloud model in their equilibrium chemistry climate model. PACT (M. Agúndez 2025) incorporates disequilibrium chemistry in gas giants as well as the ability to model secondary atmospheres of warm terrestrial planets without condensibles. coolTLUSTY (I. Hubeny & T. Lanz 1995; D. Sudarsky et al. 2005; A. Burrows et al. 2008; B. Lacy & A. Burrows 2023) can generate cloudy one-dimensional RCE models with both equilibrium and disequilibrium chemistry.

[13] NSF Graduate Research Fellow.
[14] 51 Pegasi b Fellow.

Our code, PICASO, traces its origins to the Fortran-based EGP model, which has been extensively used to study solar system atmospheres (C. P. McKay et al. 1989; M. S. Marley & C. P. McKay 1999), irradiated exoplanets (J. J. Fortney et al. 2005, 2007, 2008, 2020; M. S. Marley et al. 2012; C. V. Morley et al. 2015, 2017), and brown dwarfs (M. S. Marley et al. 1996, 2021; T. Karalidi et al. 2021; Z. Zhang et al. 2021). In an effort to make EGP more accessible to the broader community, PICASO started to be developed as its open-source, Python counterpart. PICASO 1.0 (N. E. Batalha et al. 2019) focused on computing reflected-light spectra. PICASO 2.0 expanded the package to include thermal emission and transmission spectra, as well as three-dimensional calculations in both reflection and emission (D. J. Adams et al. 2022). PICASO 3.0 (S. Mukherjee et al. 2023) incorporated the RCE modeling infrastructure at the core of EGP, enabling forward models of clear atmospheres for brown dwarfs and irradiated planets in chemical equilibrium and disequilibrium. PICASO 3.0 has been used to generate the Sonora Elf Owl grid of models (S. Mukherjee et al. 2024; N. F. Wogan et al. 2025b), which is part of the Sonora family of models that includes Sonora Bobcat (M. S. Marley et al. 2021), Sonora Cholla (T. Karalidi et al. 2021), Sonora Diamondback (C. V. Morley et al. 2024), and Sonora Red Diamondback (C. E. Davis et al. 2025), as well as in numerous other studies of brown dwarfs and exoplanets (e.g., L. Alderson et al. 2023, 2024; S. A. Beiler et al. 2023; B. E. Miles et al. 2023; Z. Rustamkulov et al. 2023; D. Powell et al. 2024; T. Konings et al. 2025). PICASO 3.0 also introduced support for computing thermal (N. Robbins-Blanch et al. 2022) and reflected-light phase curves (C. D. Hamill et al. 2024).

Through all these developments in PICASO, there were still some key features in EGP that were not incorporated. The first major functionality is fully self-consistent climate-cloud coupling. Clouds are a dominant source of opacity in many brown dwarf and exoplanet atmospheres, influencing the thermal structure, emergent spectra, and evolution. Accurate forward modeling therefore requires a realistic treatment of cloud formation, sedimentation, and radiative feedback. EGP is able to self-consistently model cloudy substellar atmospheres (C. V. Morley et al. 2012, 2014a, 2014b, 2024) using the A. S. Ackerman & M. S. Marley (2001) parameterization, but this capability was not publicly available. For the first time, PICASO 4.0 makes this cloud-climate coupling capability fully open source by integrating Virga (N. E. Batalha et al. 2026), the Python implementation of the A. S. Ackerman & M. S. Marley (2001) model, into the climate solver.

In addition to clouds, chemical disequilibrium has emerged as a critical component for accurately modeling the complex processes in substellar atmospheres. Strong vertical mixing can drive large departures from equilibrium abundances, particularly for species such as CO, $CO_2$, $CH_4$, $NH_3$, and $PH_3$, as demonstrated in recent studies (T. Karalidi et al. 2021; S. Mukherjee et al. 2022, 2025a). While PICASO 3.0 already supported disequilibrium chemistry, PICASO 4.0 extends the current capability by adding more flexible disequilibrium options and fully integrating with photochem (N. F. Wogan et al. 2025a) to generate radiative-convective-photochemical-equilibrium (RCPE) models. This development, in fact, extends the capabilities of PICASO 4.0 beyond EGP, allowing for a wider range of chemical pathways, including photochemically produced species, to be modeled self-consistently.

Finally, EGP also supports the injection of additional heating or energy into the atmosphere. Recent observations of isolated brown dwarfs, WISE 1935 (J. K. Faherty et al. 2024; G. Suárez et al. 2025) and SIMP 0136 (E. Nasedkin et al. 2025), have found thermal inversions in their upper atmospheres. The mechanism responsible for this heating remains under investigation. To enable forward modeling of such phenomena, PICASO 4.0 introduces a parameterized energy-injection module, originally developed within EGP, that allows users to deposit additional energy into the atmosphere. The PICASO implementation is more flexible than the EGP implementation, allowing custom user-supplied energy-injection profiles as well. This feature can approximate processes such as gravity-wave dissipation or magnetically driven heating mechanisms that are not typically captured in standard radiative-convective models.

These major updates in PICASO 4.0 have already been used, while still under development, in several studies (D. C. Bardalez Gagliuffi et al. 2025; C. Beichman et al. 2025; K. A. Crotts et al. 2025; S. Mukherjee et al. 2025b; S. P. Schmidt et al. 2025; A. Sanghi et al 2026). Here, we present the official release of PICASO 4.0 and its expanded functionality that builds on the foundation of the climate model introduced in PICASO 3.0. In Section 2, we describe the new capabilities and their implementation. Section 3 presents benchmark tests demonstrating agreement with previously published models. Section 4 outlines planned future developments, and Section 5 summarizes the PICASO 4.0 update.

## 2. Overview of Model Updates

S. Mukherjee et al. (2023) provided an overview of the core of the climate model, including the radiative transfer and opacity calculations, along with the framework to reach RCE. In the following sections, we elaborate on parts of the climate solver that are most pertinent to the new updates and functionalities. We focus on describing the updates from PICASO 3.0, the available options for each functionality, and how users can configure the associated parameters. Updates and new functionalities include:

1. New equilibrium chemistry tables that expand the grid to 2121 $P$, $T$ points over a wider range of atmospheric pressures (1 μbar to $10^4$ bar) and temperatures (75–6000 K) with additional tabulated outputs for the abundances for C, O, Mg, $Mg^+$, Si, $Fe^+$, Ti, $Ti^+$, and $C^+$.
2. New flexibilities for chemical disequilibrium, including rainout chemistry, cold trapping volatile species, and the complete removal of $PH_3$.
3. Integration with photochem to include the effects of photochemistry and kinetics.
4. Coupling Virga to generate self-consistent cloudy climate models.
5. The ability to change the cloud coverage fraction for different levels of patchy clouds.
6. Updated opacities for $CH_4$, $NH_3$, and inclusion of $SO_2$. This also includes more flexible methods to use the opacities for the climate models.

7. New correlated-$k$ tables for each of the 2121 chemistry grid points and individual molecules.
8. Energy injection into the atmosphere in a parameterized manner.

The above list will the major focus on of the manuscript text. However, we point out that there are also logistical changes that improve ease and usability of the code, which include:

1. New reference data structure to help the organization of backend data (opacities, `Virga` data, Sonora models, stellar grids) (✎ see “Reference Data”).
2. New get data helper (`get_data`), which will help users auto download and organize their reference data (✎ see “Quickstart for Researchers”).
3. New environment checker/visualization, which will help users to ensure they have correctly installed the data and code (✎ see “Run PICASO Environment Checker”).
4. New `picaso-lite`, which provides a quick start installation process and requires little drive space for pedagogical purposes (✎ see “Quickstart for Students and Learning”).
5. New tutorial organization to help the user to navigate the documentation.

### 2.1. Climate

Figure 1 shows the `PICASO 4.0` climate workflow, which extends the `PICASO 3.0` architecture with several new capabilities and a simplified workflow. Users supply the basic atmospheric parameters (effective temperature, surface gravity, metallicity, C/O ratio) along with initial guesses for the pressure–temperature profile, the radiative-convective boundary, and desired heat redistribution. A chemical treatment must also be selected between equilibrium chemistry and disequilibrium chemistry, with the additional option of photochemistry. For disequilibrium models, the user chooses either a self-consistent $K_{zz}$ derived from mixing-length theory (P. J. Gierasch & B. J. Conrath 1985; S. Mukherjee et al. 2022) or a constant, user-defined value. Finally, if desired, the user selects inputs related to clouds (via `Virga`) and energy injection. Following the user’s input, a workflow is triggered according to these specifications. The core framework of this climate model follows the following steps (Figure 1):

1. Read in the initial pressure–temperature ($T(P)$) profile, chemical abundance profile, and opacity.
2. Compute $K_{zz}$ (if self-consistent option requested, this incurs an extra call to retrieve net fluxes) and apply selected chemical treatments (i.e., disequilibrium chemistry quenching, photochemistry, cold trapping, rainout).
3. Compute clouds using `Virga` if requested.
4. Update total atmospheric layer optical depths if photochemistry, disequilibrium chemistry, or clouds are enabled.
5. Perform radiative transfer calculations and update the net fluxes in the atmosphere.
6. Adjust the $T(P)$ profile and convective zone.
7. Check for convergence: if converged, generate the outputs; otherwise, repeat all the steps above until convergence is reached.

We dedicate a subsection to each of these major steps in the workflow. We begin by describing the initial chemistry of the model, which is drawn from new precomputed chemical tables (Section 2.2.1). We then introduce additional chemical approximations, such as rainout and cold trapping in Section 2.2.2; these modules are optional, and the core workflow proceeds regardless of whether they are enabled. The remaining optional steps in the workflow, shown in dashed boxes in Figure 1, include photochemistry (Section 2.3) and cloud modeling (Section 2.4). We follow with our updated opacities and new opacity formats in Section 2.5. Finally, we describe the new parameterized energy-injection module (Section 2.6).

#### 2.1.1. Solving for Radiative-convective Equilibrium

Most aspects of the RCE solver were previously described in S. Mukherjee et al. (2023). One component that required revision is the maximum allowable temperature perturbation, $dT$, applied to each layer during iterations solving for the $T(P)$ profile. In version 3.0, $dT$ grew without limits, which substantially reduced runtime but introduced convergence problems, particularly for cold ($\lesssim$400 K) atmospheres and for models that included $H_2O$ clouds. In version 4.0, we impose a cap on this growth by restricting `step_max`, the variable for $dT$, to 0.05% of the layer temperature per iteration, similar to the 0.01% limit adopted in `EGP`. Although this adjustment increases the computational time, it yields smaller, better-controlled temperature updates and produces $T(P)$ profiles that converge more reliably across the full parameter space. We tested several alternative limits but ultimately found that the 0.05% cap provides the most consistent convergence while remaining more efficient than more restrictive choices.

Here, for completeness, we also describe the radiative-convective boundary solver. When initiating a climate model, the user provides the pressure–temperature profile and an initial guess for the radiative-convective boundary (`nstr_upper`). This is the top layer of the initial convective zone. For each iteration of the $T(P)$ profile, we check the flux energy balance in each layer and whether the profile is unstable against convection. Starting from the layer defined by `nstr_upper`, we allow the convective zone to grow upward in the atmosphere, including the possibility of a detached convective zone above. For this reason, it is always recommended to start with a deeper guess for the radiative-convective boundary in `PICASO`.

Figure 2 shows the iterative steps toward solving for RCE. In this example, there are 90 layers in the atmosphere, and we start with the convective zone in layer 90. From the $T(P)$ profile, the lapse rate ($d\ln T/d\ln P$) is computed for every layer of the atmosphere. We then compare these lapse rates against H/He adiabatic lapse rates from a precomputed grid (M. S. Marley et al. 2021) to determine whether each layer should be convective. For each layer, the grid is interpolated to the local pressure–temperature conditions to obtain the corresponding adiabatic lapse rate. If the ratio of the lapse rate of the atmospheric profile and the local adiabatic lapse rate in layer 89 is superadiabatic ($>$1), then the layer is forced to follow the local adiabatic lapse rate and the convective zone grows up one layer. If the ratio between the lapse rate and the local adiabatic is greater than 1.8, the convective zone grows by two layers. This process repeats and the convective zone continues to grow until a layer is stable against convection, and energy will be transported via radiation. This layer is the final radiative-convective boundary.

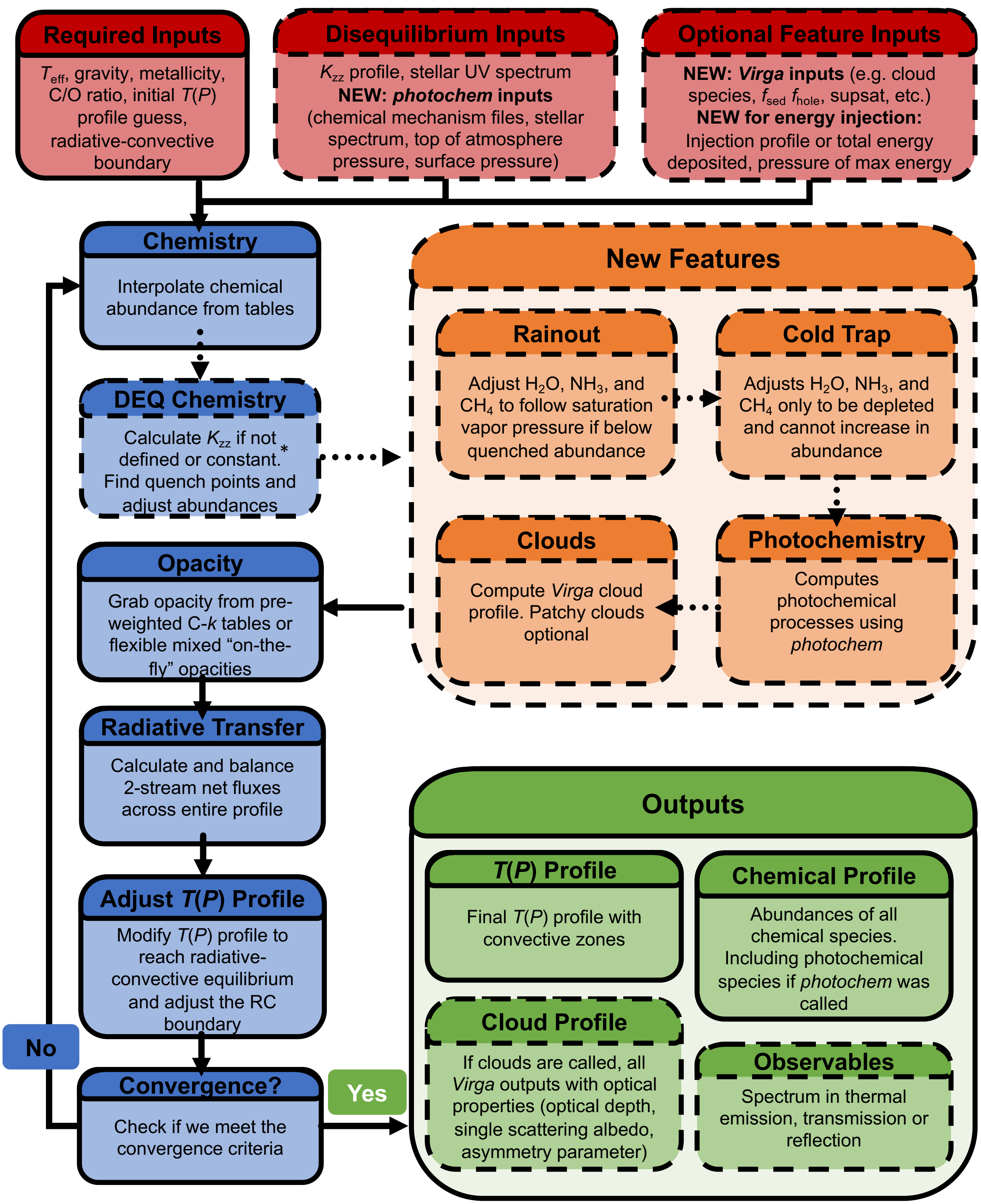

**Figure 1.** Diagram showing the climate modeling workflow of `PICASO 4.0`. Dashed boxes represent optional steps or features. Dotted arrows show the continuation of the workflow even through optional processes that are not required. Red boxes show the different inputs, with new required inputs for `photochem`, `Virga`, and energy injection. Additional input options are also shown in the GitHub tutorials. The blue boxes represent the core of the climate model computational modules. The new features are highlighted in the orange boxes and, finally, the green boxes outline the outputs from a climate model. *Note: when calculating $K_{zz}$, this also calls for a radiative transfer calculation to retrieve the fluxes across the atmospheric profile.

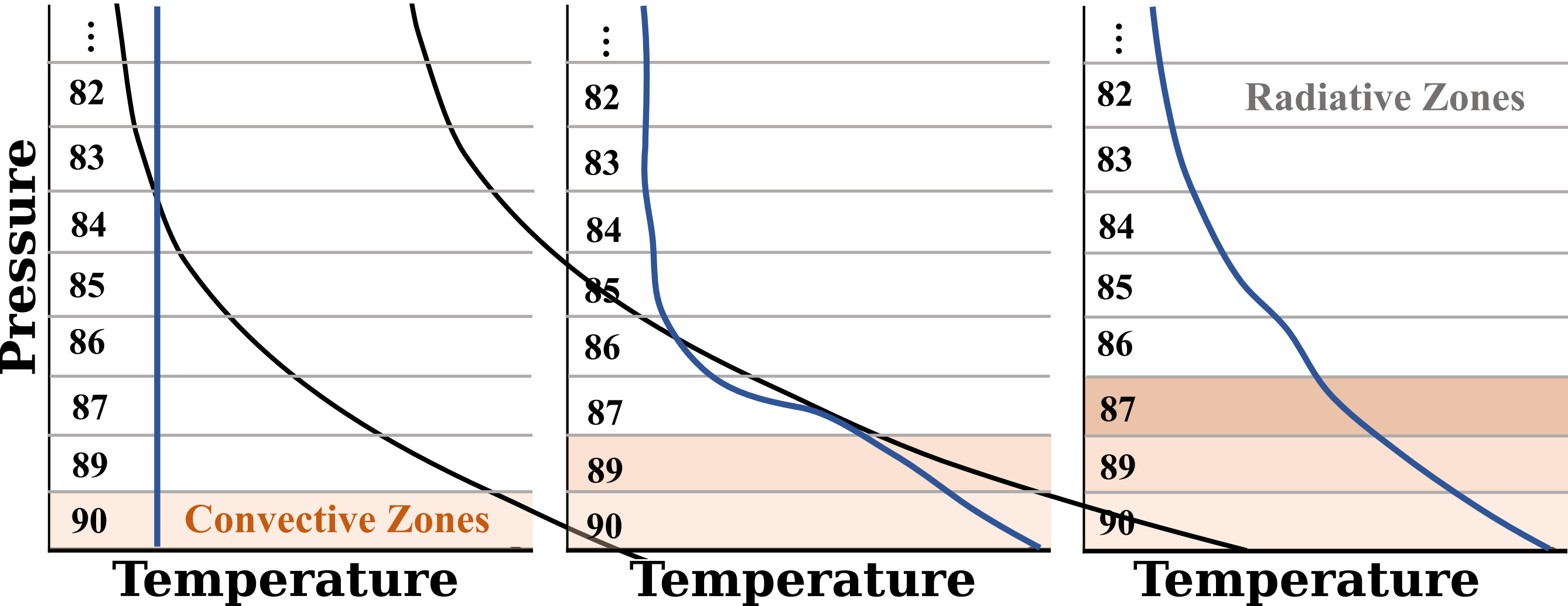

**Figure 2.** A diagram showing how PICASO solves for the radiative-convective boundary in a 90-layer climate model. Starting from the bottom of the atmosphere, the lapse rate for the climate profile is compared to the adiabatic lapse rate, and if it is superadiabatic, that layer is fixed to follow the adiabat and the convective zone grows up. This process occurs iteratively (left to right) as the climate model marches toward RCE.

It is important to note that in PICASO, a single detached convective zone is allowed. While some layers in the atmosphere above the deep convective zone may become stable against convection, regions of higher opacity further up in the atmosphere may still be present. If these layers become unstable against convection, just like the deep convective zone, they will be forced to follow the local adiabatic lapse rate. While a third or fourth detached convective zone could be possible in some specific cases (C. V. Morley et al. 2012), this is not currently implemented in the PICASO radiative-convective zone solver.

Note, our convective zone solver has largely not been modified since the original scheme was developed in M. S. Marley & C. P. McKay (1999) and C. P. McKay et al. (1989) for Uranus and Titan. Other convective zone solvers, for example the one in N. F. Wogan et al. (2025a), allow at each iterative step every layer that exceeds the adiabat to be convective. In this method, no initial guess needs to be provided. The drawback to this method is that it requires the initial $T(P)$ guess to be quite close to the final solution. If not, instabilities or nonconvergence may occur. The drawback to the method used in PICASO is that convergence can depend on the choice of initial conditions, particularly when the initial guess for the radiative-convective boundary is placed too high in altitude relative to where the final radiative convective boundary is meant to be. We always recommend initializing the radiative-convective boundary near the deepest layer of the profile. In future versions of PICASO, we will explore better radiative convective boundary solver methods. Here, however, we default to the original legacy method, which is largely reliable for the H/He atmospheres with one or two convective zones explored here.

## 2.2. Chemistry

### 2.2.1. Chemical Equilibrium Tables

The chemical equilibrium abundances tables were calculated following the general approach of M. S. Marley et al. (2021) using the easychem implementation (E. Lei & P. Mollière 2025) of the NASA CEA Gibbs minimization code (S. Gordon & B. J. McBride 1994). The current grid has been expanded from 1060 to 2121 $P$, $T$ grid points that cover a wide range of atmospheric pressures (1 μbar–$10^4$ bar) and temperatures (75–6000 K).

The equilibrium abundance grids (with results reported for key constituents that are included in the opacity calculations: e-, $H_2$, H, H+, H-, $H_2^-$, $H_2^+$, $H_3^+$, He, $H_2O$, $CH_4$, CO, $NH_3$, $N_2$, $PH_3$, $H_2S$, TiO, VO, Fe, FeH, CrH, Na, K, Rb, Cs, $CO_2$, HCN, $C_2H_2$, $C_2H_4$, $C_2H_6$, SiO, MgH, COS, Li, LiOH, LiH, LiCl, OH, $Li^+$, LiF, C, O, Mg, $Mg^+$, Si, $Fe^+$, Ti, $Ti^+$, and $C^+$) were calculated over a range of metallicities (−2.0 dex to +2.0 dex) and C/O element abundance ratios (0.25–2.0 times the solar C/O abundance ratio of C/O = 0.549) using protosolar element abundances from K. Lodders (2020). We note that our opacity list is not this exhaustive. Users should take care to ensure their species of interest also have corresponding opacities if they are interested in spectral detectability. The updated abundance tables retain equilibrium grids with C/O = 0.46 to allow for comparison with previous abundance tables (e.g., M. S. Marley et al. 2021) based on protosolar element abundances from K. Lodders (2010). For a given metallicity, variations in the C/O ratio were computed while keeping the C+O abundance constant, so that the total heavy-element abundance characterized by [M/H] remains constant.

As in previous models (e.g., see M. S. Marley et al. 2021), thermochemical data for most species are drawn from L. V. Gurvich et al. (1989, 1991, 1994), M. W. Chase (1998), A. Burcat & B. Ruscic (2005), and R. A. Robie & B. S. Hemingway (1995), together with updated low-temperature heat capacity calculations for key molecular species. Thermodynamic data for $P_4O_6$ in the updated model were taken from the Gurvich database (in place of the NIST/JANAF database; e.g., see S. Borunov et al. 1995; C. Visscher 2020; W. Bains et al. 2023), with $NH_4H_2PO_4$ included as the primary low-temperature condensate for phosphorus (C. V. Morley et al. 2018). For all condensates considered in the model, the condensation condition is defined by vapor pressure saturation of cloud-forming species (e.g., see N. E. Batalha et al. 2026). The equilibrium abundance of each cloud-forming species at higher altitudes (i.e., above the cloud) is therefore determined by its temperature-dependent vapor pressure above the condensate, effectively consistent with the "rainout" assumption for cloud formation in substellar atmospheres.

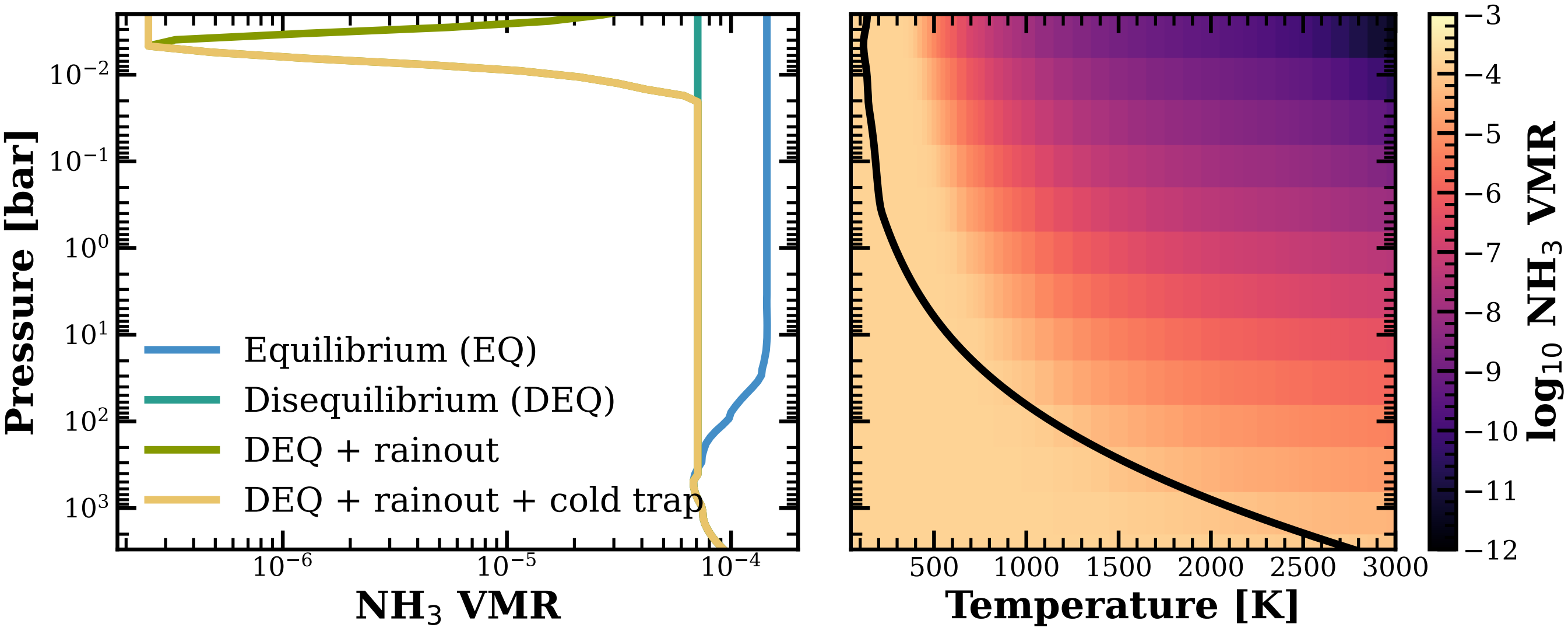

**Figure 3.** Left: Volume mixing ratio profiles of $NH_3$ in a 250 K brown dwarf with $\log(g) = 4.0$ showing the different chemistry options in PICASO: chemical equilibrium (blue), chemical disequilibrium (turquoise), disequilibrium with volatile rainout (`vol_rainout`, green), and disequilibrium with volatile rainout and "cold trapping" (`cold_trap`, yellow). Right: The brown dwarf $T(P)$ profile (black) overlaid on the $NH_3$ abundances from the updated 2121-point equilibrium chemistry table with $[\mathrm{M/H}] = +0.0$ and $\mathrm{C/O} = 0.55$.[15]

### 2.2.2. New Chemistry Approximations

There are three deviations from chemical equilibrium that we account for without more complex models. For example, we can adjust the rainout of certain species in cases where we have the condensation curves readily available. In total, we have three new options. The first option is only applicable for models in chemical disequilibrium while options 2 and 3 apply to both equilibrium and disequilibrium models:

1. `vol_rainout` (bool): This option enforces the rainout of $H_2O$, $NH_3$, and $CH_4$, even in disequilibrium chemistry runs. These three species are, by default, the only three included in our rainout scheme. However, any species that has an available condensation curve in `Virga` could be considered as a rainout species. In disequilibrium models, these species are still quenched and maintain constant abundances above the quench level. However, if the local saturation vapor pressure falls below the quenched abundance, the abundance profile will then follow the saturation vapor pressure curve.
2. `cold_trap` (bool): The abundances of volatile species such as $H_2O$, $NH_3$, and $CH_4$ are not allowed to increase once rainout begins to simulate the effect of cold trapping. This ensures that, after condensation starts, the vapor-phase abundance of a species cannot increase again at higher altitudes. Since the saturation vapor pressure depends on temperature, such increases would occur if the atmosphere warms up above the condensation layer.
3. `no_ph3` (bool): This option removes $PH_3$ entirely from the atmosphere. It was motivated by the absence of $PH_3$ detections in cold brown dwarfs, with the exception of a recent detection at the one part per billion level in WISE 0855 (M. J. Rowland et al. 2024) and $1 \times 10^{-7}$ in Wolf 1130C (A. J. Burgasser et al. 2025).

Figure 3 (left) shows abundance profiles of $NH_3$ for a 250 K brown dwarf under four scenarios: chemical equilibrium, chemical disequilibrium, disequilibrium with volatile rainout (`vol_rainout`), and disequilibrium with volatile rainout and "cold trapping" (`cold_trap`). The disequilibrium case (turquoise) uses the quenching prescription adopted in Sonora Elf Owl, where abundances are fixed above the quench level. In contrast, the chemical equilibrium case (blue) follows the equilibrium table (right), showing a smooth increase in $NH_3$ with decreasing temperature. When cold enough, the `vol_rainout` option (green) captures the rainout of $NH_3$ in disequilibrium by enforcing condensation along the saturation vapor pressure curve. However, in regions where the upper-atmosphere temperature rises above the condensation curve, the $NH_3$ abundance will increase. The `cold_trap` option (yellow) prevents this by holding the abundance constant at the value from the colder layer below.

Additionally, we update the treatment of $CO_2$ in chemical disequilibrium. S. A. Beiler et al. (2024) identified an underprediction of $CO_2$ in atmospheric profiles when Sonora Elf Owl models (S. Mukherjee et al. 2024) were compared with JWST observations of cool brown dwarfs. N. F. Wogan et al. (2025b) released Sonora Elf Owl 2.0, which applies an empirical correction to the $CO_2$ abundance and provides an in-depth description of the adjustments made. Now, `PICASO 4.0` implements the self-consistent adjustment of $CO_2$ following the quenching prescription in K. J. Zahnle & M. S. Marley (2014).

### 2.3. Photochemistry

`PICASO` can optionally self-consistently determine composition with the one-dimensional photochemical model in the `Photochem` software package (N. F. Wogan et al. 2025a). This functionality was first introduced to `PICASO` in

[15] Chemistry Tutorial

S. Mukherjee et al. (2025a) (though not officially open-sourced), and here we give a brief overview of the feature to demonstrate how it relates to the other chemistry treatments. Given an input $P$–$T$–$K_{zz}$ profile within step 2 of the `PICASO` climate solver (Section 2.1), `Photochem` solves for the steady-state of a system of partial differential equations describing how atmospheric gases are influenced by chemical reactions, condensation/evaporation, and vertical transport (Equation (1) in N. F. Wogan et al. 2025a). The bottom of the photochemistry model domain is fixed to chemical equilibrium computed with the equilibrium solver in `Photochem`, and the top of the atmosphere has zero-flux boundary conditions for all species.

When generating a `PICASO`-`Photochem` coupled model, radiative transfer is solved independently within each framework, but over complementary wavelength regimes. `PICASO` computes radiative transfer for solar and infrared fluxes of a planet, while `Photochem` solves UV radiative transfer to compute photolysis rates using an extensive set of photoabsorption cross sections for dozens of species. Although this introduces minor inconsistencies at overlapping wavelengths (e.g., in the visible), the dominant photolysis occurs at $\lambda \lesssim 250$ nm, where `PICASO` does not include the relevant opacities.

Any simulation that leverages `Photochem`, by default, assumes chemical equilibrium for species that `Photochem` does not track. `Photochem` comes with a network of ∼610 reversible chemical reactions and ∼95 photolysis reactions involving 98 species composed of H, He, N, O, C, S and Cl, although it is straightforward to use other networks. Consequently, the code does not predict the composition of species containing Fe, Si, Na, etc., which are often important opacity sources in the deep atmosphere of a gas-rich planet. So, in practice, the code first computes chemical equilibrium (Section 2.2.1) and then runs the photochemical model to a steady-state, updating the equilibrium result with disequilibrium profiles only for species considered by `Photochem`.

### 2.4. Clouds

To enable fully self-consistent coupling between the climate and cloud model, we integrate `Virga` (N. Batalha et al. 2020; N. E. Batalha et al. 2026; S. E. Moran et al. 2025) into the `PICASO` RCE framework. `Virga` is a Python implementation of the widely used `EddySed` model originally described by A. S. Ackerman & M. S. Marley (2001), and adopted in numerous studies (M. S. Marley et al. 2010; C. V. Morley et al. 2012, 2014a, 2014b, 2015; A. J. Skemer et al. 2016) as well as in grid models such as D. Saumon & M. S. Marley (2008) and the Sonora Diamondback models (C. V. Morley et al. 2024). This parameterized cloud model balances the upward mixing of vapor with the downward settling of cloud particles,

$$-K_{zz}\frac{\partial q_t}{\partial z} = f_{\rm sed}\, w_* q_c, \qquad (1)$$

where $K_{zz}$ is the eddy diffusion coefficient, $q_t$ is the total mass mixing ratio of the cloud condensate + vapor, $z$ is the altitude, $f_{\rm sed}$ is the sedimentation efficiency parameter, $w_*$ is the mean upward velocity, and $q_c$ is the mass mixing ratio of only the cloud condensate. Microphysical processes such as nucleation and coagulation of cloud formation are not considered; instead, all vapor in excess of the saturation vapor pressure condenses into a cloud. The cloud particle size distribution in `Virga` follows a lognormal distribution. The impacts of these assumptions on the climate thermal structure, convergence behavior, and observables are explored in J. Mang et al. (2022, 2024).

To initiate a cloud self-consistently within the `PICASO` climate model, users provide the same inputs required for a standalone `Virga` calculation, including the condensate species and $f_{\rm sed}$. Larger values of $f_{\rm sed}$ produce vertically thinner, optically thinner clouds with larger particles, whereas smaller values yield more vertically extended, optically thicker clouds. A variable $f_{\rm sed}$ prescription introduced into `Virga` (C. M. Rooney et al. 2022) is also available to be used in `PICASO 4.0` when the user sets `param = "exp,"`

$$f_{\rm sed}(z) = \alpha \exp\left(\frac{z - z_{\rm T}}{6\beta H_0}\right) + \epsilon, \qquad (2)$$

where $\alpha$ is a constant of proportionality which in application corresponds to the $f_{\rm sed}$ value at the top of the atmosphere, $z$ is the altitude, $z_{\rm T}$ is the altitude of the top layer of the atmosphere, $\beta$ is the scaling parameter which defines the $f_{\rm sed}$ falloff rate, $H_0$ is the scale height, and $\epsilon$ is the minimum allowable $f_{\rm sed}$ value.

Users may additionally specify a supersaturation factor via the optional argument `supsat`, which increases the saturation vapor pressure and suppresses cloud formation,

$$\text{supsat} = \frac{q_v(z - \Delta z) - q_c(z)}{q_s(z)} - 1, \qquad (3)$$

where $q_v$ is the vapor mass mixing ratio, $q_c$ is the condensate mass mixing ratio, and $q_s$ is the saturation vapor mass mixing ratio. By default, `supsat = 0`, increasing the value reduces the amount of cloud formation. Guidance on choosing these parameters to match microphysical cloud profiles in cold substellar atmospheres is provided in J. Mang et al. (2024). The cloud outputs from the `PICASO` climate model include the data frame containing all relevant cloud properties (e.g., optical depth, single scattering albedo, asymmetry parameter, particle size distribution) calculated with `Virga` and formatted for direct use with its plotting tools to generate diagnostic plots. Additional details on the `Virga` workflow, optical properties, and condensate treatment are provided in N. E. Batalha et al. (2026).

Two important methodological choices were made to ensure stable cloud-climate coupling and that the climate model achieves RCE. The first choice is the treatment of $K_{zz}$ in the atmosphere. $K_{zz}$ remains a major source of uncertainty in the atmospheric modeling of brown dwarfs and exoplanets (S. Mukherjee et al. 2022). Though `PICASO 4.0` does allow for constant $K_{zz}$ profiles, those constant profiles are only used for disequilibrium chemistry. For clouds, self-consistent $K_{zz}$ profiles are always used. In convective zones, we adopt the mixing-length theory prescription of P. J. Gierasch & B. J. Conrath (1985),

$$K_{zz} = \frac{H}{3}\left(\frac{L}{H}\right)^{4/3}\left(\frac{RF}{\mu\rho_{\rm a} c_{\rm p}}\right)^{1/3}, \qquad (4)$$

where $H$ is the scale height, $L$ is the mixing length, $R$ is the gas constant, $\mu$ is the mean molecular weight, $\rho_a$ is the density of the atmosphere, $c_p$ is the specific heat capacity at constant pressure, and $F$ is the convective heat flux throughout the atmosphere. Parameterizations in radiative zones vary widely. `PICASO 3.0` used the clear-atmosphere prescription of J. I. Moses et al. (2022), motivated by the analysis in S. Mukherjee et al. (2022),

$$K_{zz} = \frac{5 \times 10^8}{\sqrt{P_{\rm bar}}}\left(\frac{H_{1\,\rm mbar}}{620\ \rm km}\right)\left(\frac{T_{\rm eff}}{1450\ \rm K}\right)^4, \quad (5)$$

where $P_{\rm bar}$ is the pressure of the layer and $H_{1\ \rm mbar}$ is the scale height at 1 mbar. Other microphysical cloud models adopt smaller radiative-zone mixing lengths (e.g., $L = 0.1H$), which yield significantly lower $K_{zz}$ values (J. Mang et al. 2022). Because $K_{zz}$ regulates particle lofting relative to gravitational settling, very small values produce smaller grains and dramatically increase cloud column densities and optical depths, often leading to numerical instabilities. Additionally, the convergence of self-consistent atmosphere models with clouds has always been difficult since cloud profiles are extremely sensitive to small perturbations in the pressure–temperature profile. To minimize these effects, `PICASO 4.0` applies mixing-length theory estimates throughout the atmosphere and computes the radiative zone $K_{zz}$ using an average over two scale heights, smoothing abrupt layer-to-layer variations. When generating self-consistent cloudy models, $K_{zz}$ is calculated within `PICASO` following the precomputed adiabatic lapse rate described above. It is important to differentiate from how $K_{zz}$ is computed in standalone `Virga` models where the adiabatic lapse rate is estimated to be $T_i/(7/2)$, where $T_i$ is the temperature of the layer, and 7/2 is the specific heat coefficient for a permanent diatomic gas like $H_2$.

The second methodological choice made for self-consistent clouds is when to stop updating the cloud profile to ensure we reach convergence. During the climate iterations, the $T(P)$ profile changes in response to the added cloud opacity, which in turn updates the chemistry, opacities, and radiative transfer. In a fully dynamical atmosphere this cycle would continue indefinitely, but our one-dimensional static models seek a single $T(P)$ profile that satisfies RCE and reproduces the desired $T_{\rm eff}$. Once the $T(P)$ profile converges, we freeze the cloud profile, i.e., we do not perform an additional `Virga` update after the final $T(P)$ adjustment. This leaves the cloud profile one iteration behind the final thermal structure. The impact of this choice is most significant when the converged $T(P)$ profile lies close to the condensation curve of the cloud species. Water clouds, for example, are a large source of opacity that forms at ~225–275 K and can strongly affect convergence behavior. N. E. Batalha et al. (2026) also discusses the impact of this with respect to trying to reproduce the stored cloud profiles from a climate model with `Virga` using the final $T(P)$ profile. Users should be aware that these two profiles will not always match identically.

### 2.4.1. Patchy Clouds

In addition to generating fully cloudy models, the user now has the option to generate partly cloudy, or "patchy-cloud," models. This follows the method first derived in M. S. Marley et al. (2010) and further used to generate models in C. V. Morley et al. (2014a, 2014b). Patchy clouds help produce a more realistic one-dimensional atmospheric column where clouds do not homogeneously cover the entire atmosphere. In the solar system, for example, Jupiter exhibits 5 μm hot spots and banded cloud structures (B. E. Carlson et al. 1992; M. Roos-Serote et al. 2000; J. Arregi et al. 2006), while Saturn shows similarly heterogeneous cloud patterns (K. H. Baines et al. 2005). For these patchy clouds, our one-dimensional atmospheric column is split into two columns, a clear column and a cloudy column, where the fluxes from these are summed together to generate a fractionated cloud model. The total flux in each atmospheric layer is therefore controlled by the fraction of clouds to be included, $f_{\rm hole}$,

$$F_{\rm total} = f_{\rm hole}F_{\rm clear} + (1 - f_{\rm hole})F_{\rm cloudy}. \quad (6)$$

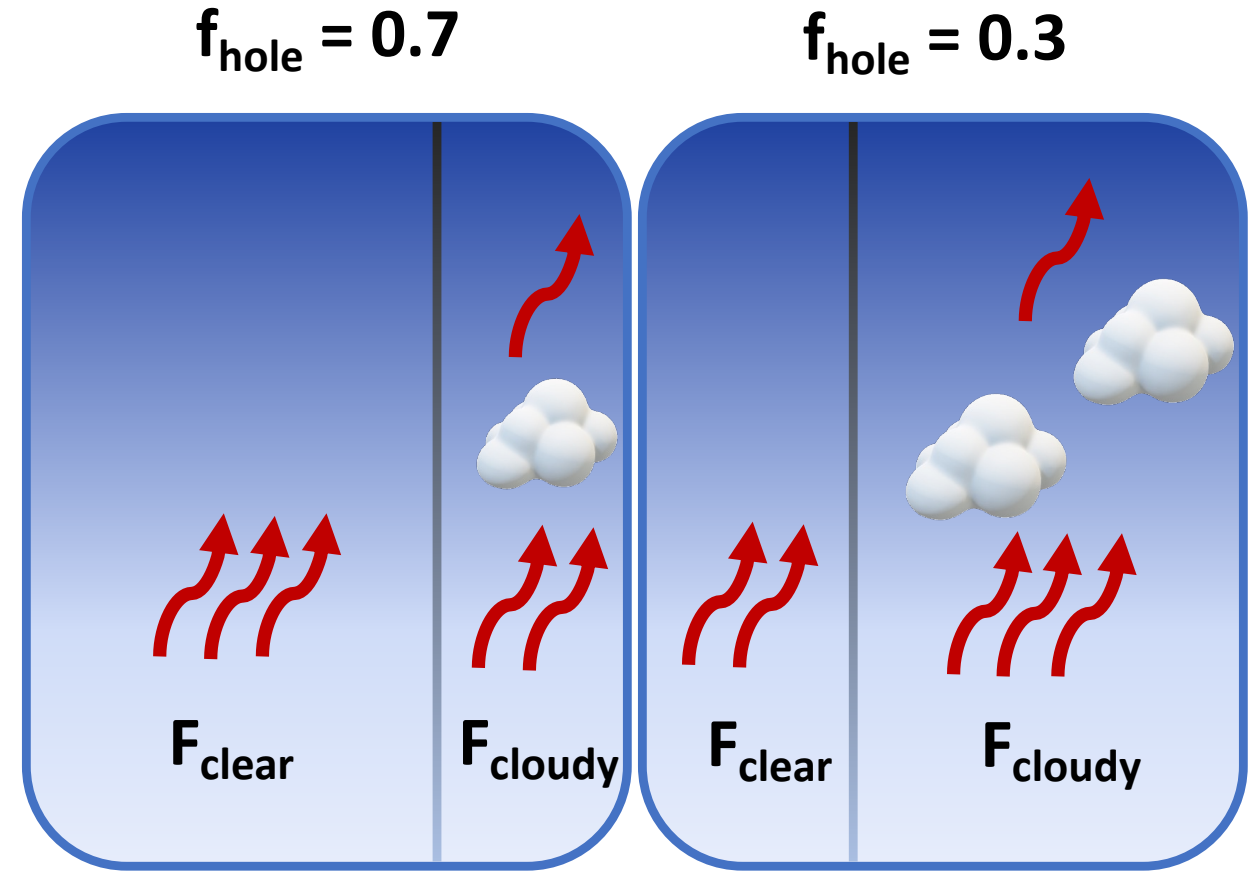

**Figure 4.** A depiction of different combinations of patchy clouds in the atmosphere tuning the parameters of `fhole`, the fraction of clear atmosphere in the model. $F_{\rm clear}$ and $F_{\rm cloudy}$ represent the total flux from the clear and cloudy column, respectively (Equation (6)).

These are the new user inputs to control the patchy clouds:

1. `do_holes` (bool): This turns on patchy clouds, or what we call "holes" in the clouds.
2. `fhole` (float): This defines the amount of cloud fractionation: 0 represents a fully cloudy model, and as the user increases the number, the model becomes less cloudy.
3. `fthin_cld` (float): This parameter increases the flexibility of the clear column, where the user is able to add a second cloud fraction. `fthin_cld = 0` is the default and keeps the clear column 100% clear. If it is increased to 0.6, then the clear column will no longer be fully clear but would be 60% cloudy.

Figure 4 depicts the atmosphere with different values of `fhole` in the atmosphere. For example, `fhole = 0.3` generates a model comprised of 30% of the clear column and 70% of the cloudy column. While a higher `fhole` value, like 0.7 represents a model that will have 70% clear atmosphere and 30% cloudy atmosphere. Future developments will allow the user more control over what cloud species, $f_{\rm sed}$ values, and fractionation to specify between the two columns.

Figure 5 compares self-consistent atmospheric models with clear, fully cloudy, and patchy-cloud scenarios (30% cloud coverage; $f_{\rm hole} = 0.7$). Relative to the clear Sonora Bobcat spectrum (black), the fully cloudy model (blue) shows a strong water-ice absorption feature between 4 and 5 μm, driven by the large cloud opacity that suppresses the outgoing flux. When the cloud coverage is reduced, the patchy-cloud spectrum (orange) exhibits intermediate flux suppression, falling naturally between the clear and fully cloudy limits.

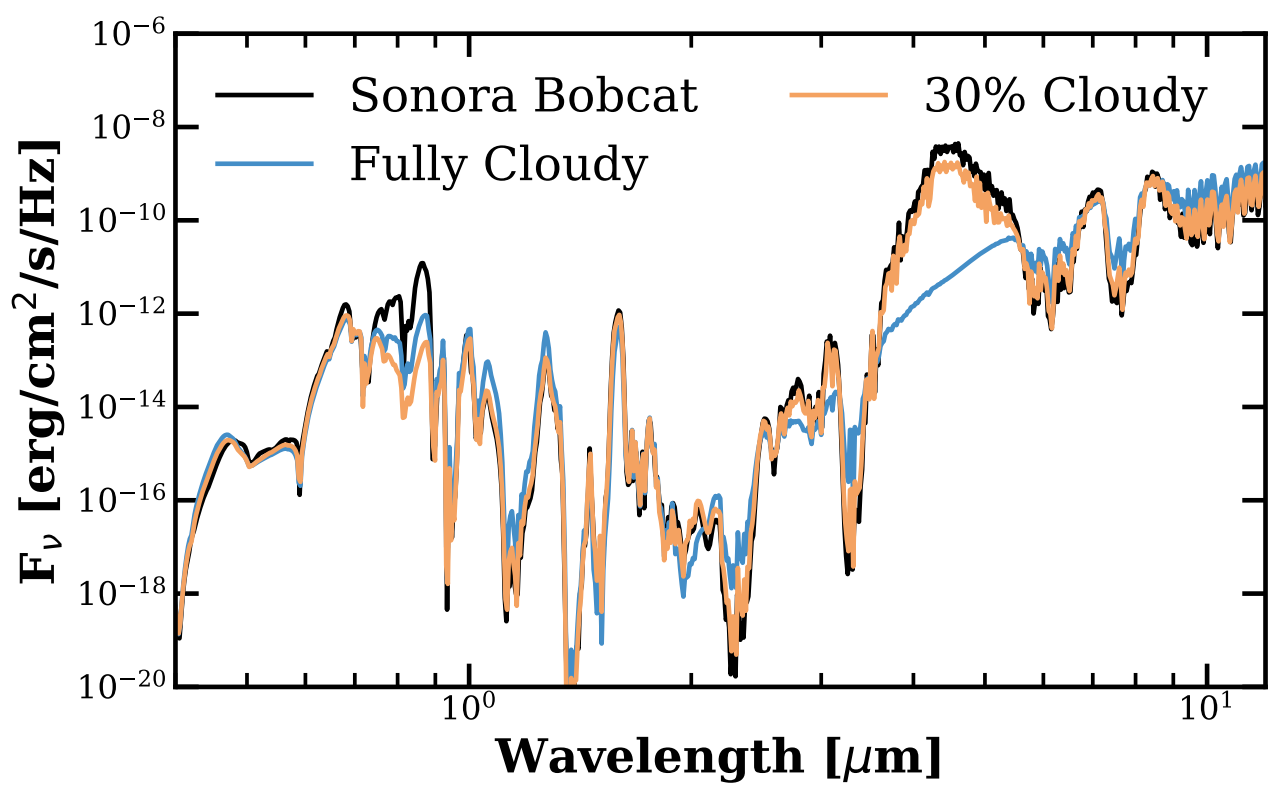

**Figure 5.** Thermal emission spectra of a clear Sonora Bobcat model (black), fully cloudy model (blue), and a 30% cloudy model (orange). These are for a 200 K brown dwarf with $\log(g) = 4$. For the cloudy models, only $H_2O$ clouds are included with $f_{sed} = 8$.[16]

### 2.5. Flexible Opacities

In `PICASO 4.0` there are three methods for incorporating opacities: resort-rebin correlated-*k* tables (`resort-rebin`), preweighted chemical equilibrium correlated-*k* tables (`preweighted`), and higher resolution ($R$ = 15,000–60,000) `resampled` cross sections. Though these three methods rely on the same backend line-by-line cross-section database, they are delivered in three separate open data repositories. In this section, we first describe the updates to the backend line-by-line cross-section database when compared to `PICASO 3` and then we describe the three different data products we supply and what they are best suited for.

For completeness here, our opacity tables are computed on a 1460 pressure–temperature grid that covers pressures $1 \times 10^{-6}$–3000 bars and temperatures 75–4000 K. They are computed on a high-resolution grid with a pressure-dependent $\Delta$-wavenumber that ranges from 0.0035 cm$^{-1}$ at $1 \times 10^{-6}$ bar and 0.5 cm$^{-1}$ at 3000 bar. We include a full wavenumber grid from 30 to 30,000 cm$^{-1}$. Our full 1460 grid can be found in this Zenodo posting (E. Gharib-Nezhad et al. 2021a) and in the PICASO reference data.[17]

In `PICASO 4.0` we update the line lists sources of $NH_3$ and $CH_4$, and add $SO_2$ (previously not included in our base opacity sets). For $CH_4$, we update our line lists from ExoMol (S. N. Yurchenko & J. Tennyson 2014) to the newer HITEMP $CH_4$ (R. J. Hargreaves et al. 2020) list. We use the methodology described in E. Gharib-Nezhad et al. (2021b) to create high-resolution cross sections on a grid of 1460 *P*–*T* points to match the standard format of our previous list. We use the wing cutoff standards described in E. S. Gharib-Nezhad et al. (2024). For $NH_3$, we update our line list source from S. N. Yurchenko et al. (2011) and J. S. Wilzewski et al. (2016) to the updated ExoMol line list from P. A. Coles et al. (2019) for $^{14}NH_3$ and S. N. Yurchenko et al. (2024) for $^{15}NH_3$. We use the methodology from K. L. Chubb et al. (2021). Unlike our treatment of all other species where we apply the standard HITRAN isotopologue weightings, we only include the main isotopologue of $CH_4$ in our correlated-*k* tables and in the publicly distributed resampled opacity database. This special treatment for $CH_4$ is motivated by recent brown dwarf analyses (B. W. P. Lew et al. 2024; M. J. Rowland et al. 2024), which

[16] Cloudy Brown Dwarf Tutorial

[17] https://github.com/natashabatalha/picaso/blob/efc4a8b514238889c55cc6523a501b92d6995c57/reference/opacities/grid1460.csv

**Table 1**
References of Gaseous Opacities Used for Calculating the Atmospheric Models and Resulting Spectra in this Work

| | |
|---|---|
| $C_2H_2$ | L. S. Rothman et al. (2013) |
| $C_2H_4$ | L. S. Rothman et al. (2013) |
| $C_2H_6$ | L. S. Rothman et al. (2013) |
| $CH_4$ | S. N. Yurchenko et al. (2013), S. N. Yurchenko & J. Tennyson (2014), C. Wenger & J. Champion (1998), A. S. Pine (1992), R. J. Hargreaves et al. (2020), computed with methods in E. Gharib-Nezhad et al. (2021b) |
| CO | L. S. Rothman et al. (2010), I. E. Gordon et al. (2017), G. Li et al. (2015) |
| $CO_2$ | X. Huang et al. (2014) |
| CrH | A. Burrows et al. (2002), computed in E. Gharib-Nezhad et al. (2021b) |
| Cs | T. Ryabchikova et al. (2015) |
| Fe | T. Ryabchikova et al. (2015), T. R. O'Brian et al. (1991), J. R. Fuhr et al. (1988), A. Bard et al. (1991), A. Bard & M. Kock (1994) |
| FeH | M. Dulick et al. (2003), with added E-A transition R. J. Hargreaves et al. (2010) |
| $H_2$ | I. E. Gordon et al. (2017) |
| $H_3^+$ | I. I. Mizus et al. (2017) |
| $H_2$–$H_2$ | D. Saumon et al. (2012) with added overtone from P. Lenzuni et al. (1991) Table 8 |
| $H_2$–He | D. Saumon et al. (2012) |
| $H_2$–$N_2$ | D. Saumon et al. (2012) |
| $H_2$–$CH_4$ | D. Saumon et al. (2012) |
| $H_2^-$ | K. Bell (1980) |
| $H^-$ bf | T. L. John (1988) |
| $H^-$ ff | K. L. Bell & K. A. Berrington (1987) |
| $H_2O$ | O. L. Polyansky et al. (2018) |
| $H_2S$ | A. A. A. Azzam et al. (2016) |
| HCN | G. J. Harris et al. (2006), R. J. Barber et al. (2014), I. E. Gordon et al. (2022) |
| LiCl | D. M. Bittner & P. F. Bernath (2018) |
| LiF | D. M. Bittner & P. F. Bernath (2018) |
| LiH | C. M. Coppola et al. (2011) |
| MgH | B. Yadin et al. (2012), E. Gharib-Nezhad et al. (2013) computed in E. Gharib-Nezhad et al. (2021b) |
| $N_2$ | L. S. Rothman et al. (2013) |
| $NH_3$ | P. A. Coles et al. (2019), S. N. Yurchenko et al. (2024), computed with methods in K. L. Chubb et al. (2021) |
| OCS | I. E. Gordon et al. (2017) |
| $PH_3$ | C. Sousa-Silva et al. (2014) |
| Rb | T. Ryabchikova et al. (2015) |
| SiO | E. J. Barton et al. (2013), E. Gharib-Nezhad et al. (2021b) |
| $SO_2$ | D. S. Underwood et al. (2016) |
| TiO | L. K. McKemmish et al. (2019) computed in E. Gharib-Nezhad et al. (2021b) |
| VO | L. K. McKemmish et al. (2016) computed in E. Gharib-Nezhad et al. (2021b) |
| Li,Na,K | T. Ryabchikova et al. (2015), F. Allard et al. (2007), N. F. Allard et al. (2007, 2016, 2019), as compiled in P. Mollière et al. (2019) |

**Note.** If where the cross section was computed is not specified, then it was compiled and computed via the framework discussed in R. S. Freedman et al. (2008, 2014).

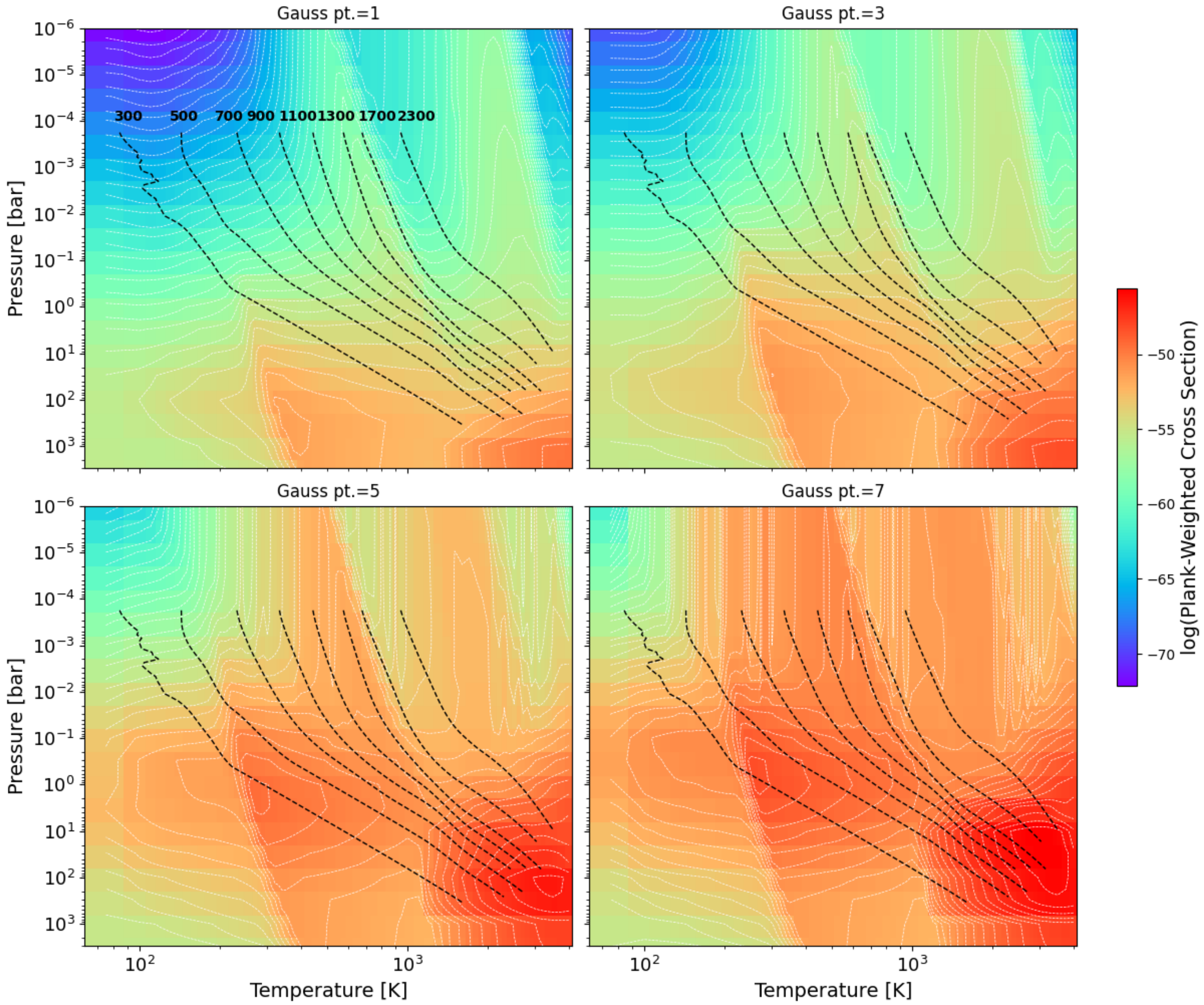

**Figure 6.** A heat map of Planck-weighted correlated-*k* tables to complement Figure 5 of `PICASO 3.0` release in order to show any major differences between opacity and chemistry updates. The four panels correspond to Gauss points 1, 3, 5, and 7 (shown in the title of each panel). The dashed black lines correspond to Sonora Bobcat pressure–temperature profiles at increasing effective temperatures for reference (as labeled in the upper left panel) (M. S. Marley et al. 2021). All major features of the correlated-*k* tables remain the same between v3 and v4.

showed that weighting $CH_4$ to $CH_3D$ using standard HITRAN isotopic ratios produced spectral features inconsistent with observations. Evolutionary models (e.g., D. S. Spiegel et al. 2011; C. V. Morley et al. 2024) indicate that sufficiently massive objects deplete their deuterium early in their lifetime through deuterium fusion, naturally suppressing the abundance of deuterated species such as $CH_3D$. Finally, for $SO_2$ we use the line list sources of D. S. Underwood et al. (2016). For completeness, a full updated table of our line list sources is included in Table 1.

Our publicly available default resampled opacity database includes all available isotopologues combined using standard isotopic weightings, with $CH_4$ being the only exception, as explained. `PICASO` supports modeling individual isotopologues (like other $CH_4$ isotopologues) when the corresponding cross sections are included in the `sqlite` opacity database. Users may generate a custom opacity database to include any different opacities following the workflow described in our opacity database tutorial. We have a longer-term effort to make additional unique opacities more easily accessible via the MAESTRO project (described in E. S. Gharib-Nezhad et al. 2024). In practice, specifying individual isotopologues is as simple as using the full isotopologue name (e.g., `12C-H4` and/or `12C-D-H4` rather than `CH4`), provided the associated cross sections are present in the database.

Given our standard 1460 $T(P)$-grid of opacities, we proceed to compute our three different opacity products, which we describe below:

1. Preweighted correlated-*k* tables (method = `pre-weighted`): Section 2.1.4 of S. Mukherjee et al. (2023) details our "double-Gauss" method for computing correlated-*k* tables (see Equation (25)). In S. Mukherjee et al. (2023), we previously did not have an open-source pipeline to do these calculations. As part of `PICASO 4.0` this is now included in pure Python in `opacity_factory.compute_ck_molecular`. To compute our

preweighted-$k$ tables on a 196 wavelength grid, we use the precomputed chemistry grid described here in Section 2.2.1, which makes our table primarily a function of metallicity and carbon-to-oxygen ratio. We include the chemistry and opacity sources of everything in Table 1. Of note, we create one complete table with TiO and VO and another table without (N. E. Batalha and J. Mang 2026a). Figure 6 shows a heat map of our preweighted correlated-$k$ tables that include both TiO and VO for solar M/H and C/O = 0.46 when they are Planck-weighted. This figure is meant to reproduce Figure 5 of the `PICASO 3` manuscript (S. Mukherjee et al. 2023). We show that our new chemistry and updates to opacities still produce all the major features of our older tables.

2. Individual molecule correlated-$k$ tables (method = `resortrebin`): Unlike the previous method, which first computes high-resolution mixed opacity tables, in this method we compute correlated-$k$ tables individually for each molecule on a 661 wavelength grid (N. E. Batalha and J. Mang 2026b). Then, the code mixes these individual molecules "on-the-fly" using the resort-rebinning scheme described in D. S. Amundsen et al. (2017). This was previously included and described in `PICASO 3.0`. Though the general methodology remains unchanged, the code itself was updated to include a flexible approach to mix any number of gases for which a table is available. Users must specify which via the `preload_gases` keyword in `opannection`.
3. Resampled opacities (method = `resampled`): This method is primarily used for only forward models (not climate). This method has also remained unchanged since the original `PICASO 1` release. All line-by-line opacities are interpolated to the same $R = 1.6 \times 10^6$ common grid and then resampled (via simple "every-other-point") to the resolution specified in the opacity database. We usually supply one grid at $R = 15{,}000$, and another at $R = 60{,}000$. We advise the use of $R = 15{,}000$ for data with $R < 300$ and $R = 60{,}000$ for data roughly at $R \sim 3000$. We have a notebook showcasing errors induced via sampling here.

### 2.6. Energy Injection

An additional artificial energy source can be injected into the climate profile to represent localized atmospheric perturbations, such as a hot spot, or to mimic temperature inversions due to unknown energy deposition (J. K. Faherty et al. 2024; E. Nasedkin et al. 2025; G. Suárez et al. 2025). To activate this feature, the user must include the `energy_injection` object. The energy injection works by monotonically increasing, layer by layer over some range of model layers, the radiative flux the atmosphere must emit in order to stay in radiative equilibrium. Two injection methods are available:

1. *Chapman function*. The Chapman function is a parameterized vertical heating profile that describes how incident flux is absorbed in an atmosphere due to additional sources of opacity. This formulation has been used to encapsulate atmospheric heating in substellar objects (M. S. Marley & C. P. McKay 1999; C. V. Morley et al. 2014a). Key inputs include the total injected energy flux in units of erg $cm^{-2}$ $s^{-1}$ (`total_energy_injection`), the pressure level at which the heating peaks in bars (`press_max_energy`), and the number of scale heights over which the Chapman profile extends (`injection_scaleheight`).
2. *User-supplied energy distribution*. This more flexible option allows the user to specify an arbitrary vertical energy-injection profile. Setting `inject_beam=True` enables the injection of a custom array `beam_profile`, which must contain layer-by-layer energy fluxes (in erg $cm^{-2}$ $s^{-1}$) and have the same length as the atmospheric grid.

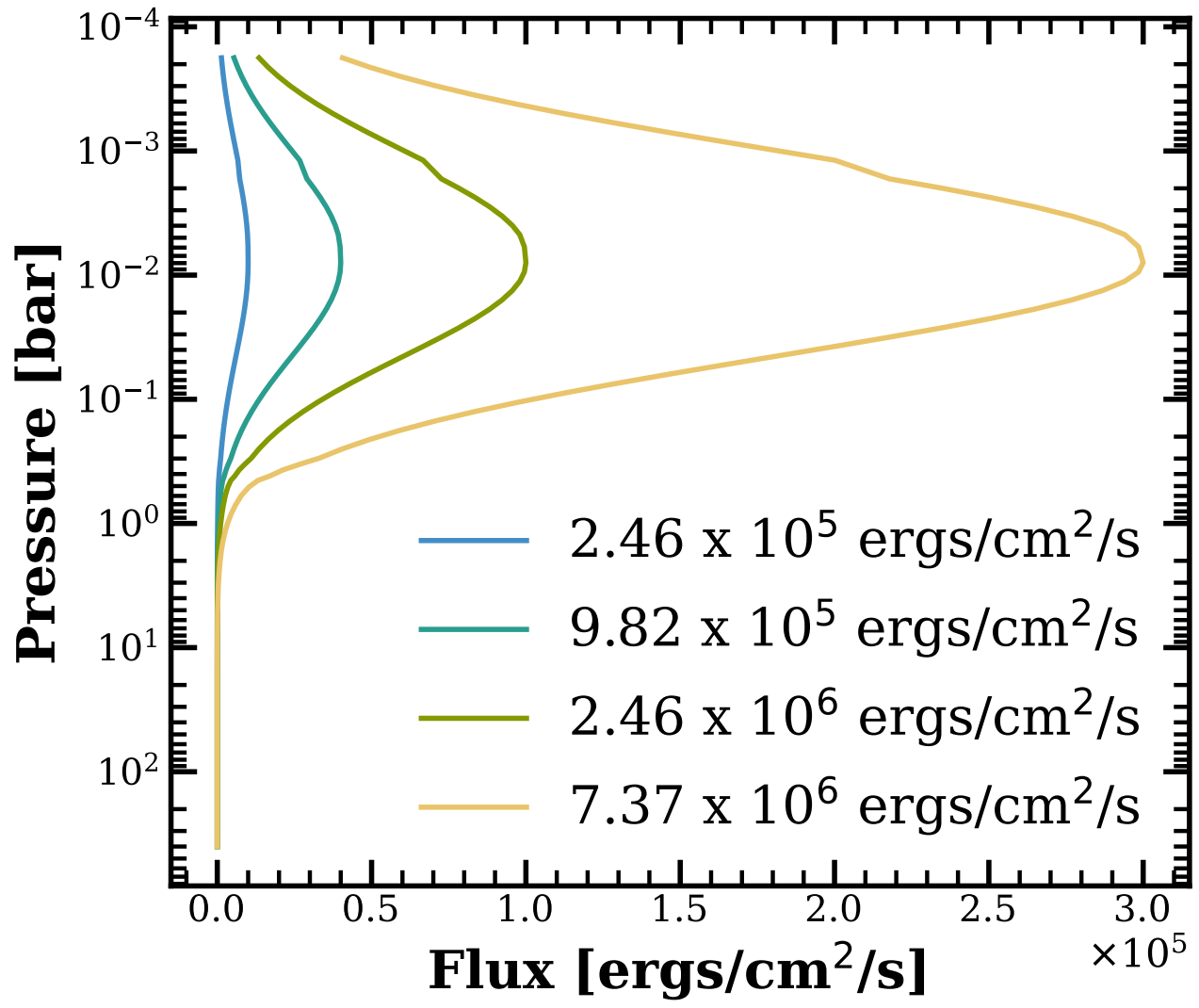

**Figure 7.** Examples of arbitrary energy deposition profiles following a Gaussian distribution. Each profile is labeled by the total amount of external energy being deposited into the atmosphere.

Figure 7 shows an example of a user-supplied energy distribution. In this case, we adopt a Gaussian profile to illustrate energy deposition across 10 atmospheric layers. During each climate iteration, the model again solves for RCE while accounting for the added energy in each layer. In the example shown, the peak energy input occurs in the 20th model layer, corresponding to a pressure of roughly $7 \times 10^{-3}$ bar. The total amount of energy injected into the atmosphere is given by the integral of the distribution across all layers. Additional tests of this feature—and its influence on the resulting thermal structure—are presented in Section 3.3.

## 3. Benchmarks

To validate the new capabilities introduced in `PICASO 4.0`, we benchmark our models against previously published results. Benchmarking ensures that the new features, cloud modeling, energy injection, and photochemistry, behave consistently with expectations from the literature. We therefore test `PICASO 4.0` across four representative scenarios: (1) comparison with the cloudy Sonora Diamondback grid to assess baseline thermal structure behavior and convergence for cloudy atmospheres, (2) reproduce the patchy-cloud profile computed in C. V. Morley et al. (2014b) and compare the thermal emission spectra, (3) compare the $T(P)$ profiles of the atmospheres with thermal inversions computed in C. V. Morley et al. (2014a) to validate the energy-injection parameterization, and (4) reproduction of published photochemical models of WASP-39 b (S.-M. Tsai et al. 2023) to demonstrate the integration of `photochem`.

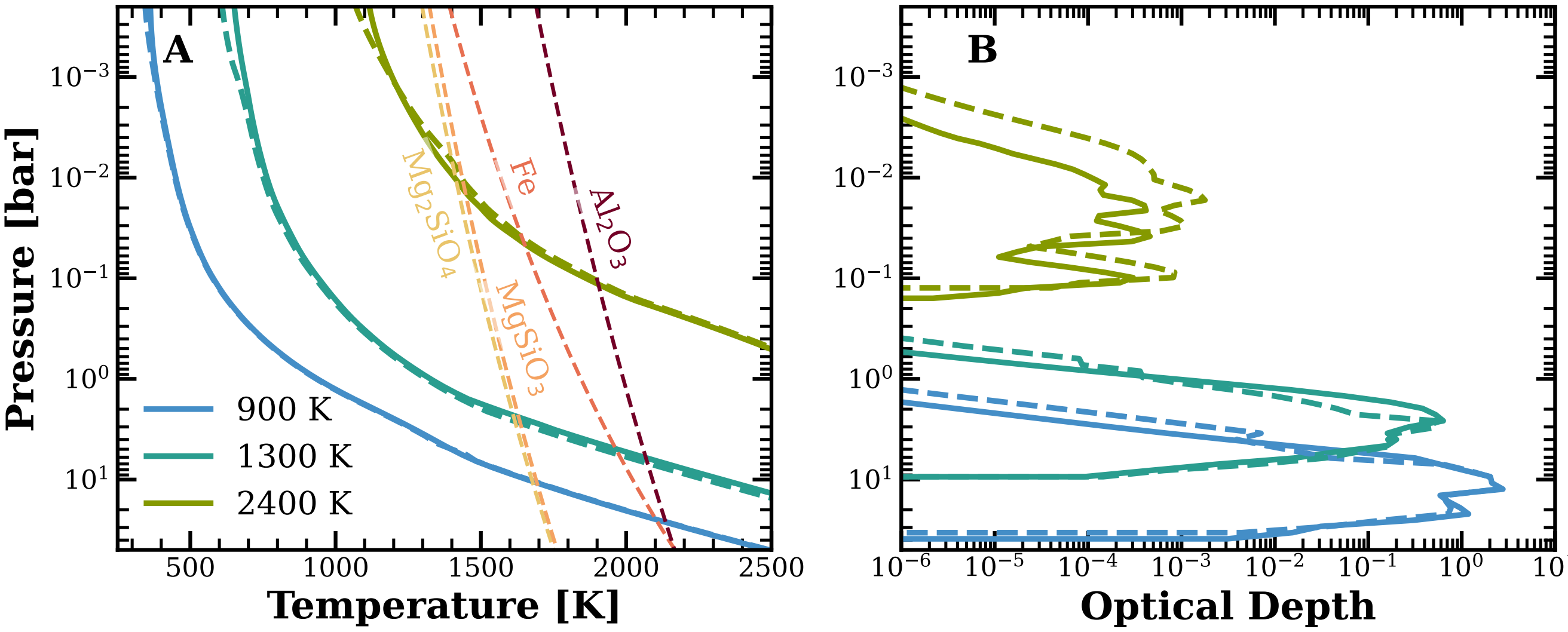

**Figure 8.** Benchmark comparison of the pressure–temperature profiles (left) and cloud optical depth profiles at 4.2 μm (right) for brown dwarfs with $T_{\mathrm{eff}} = 900$ (blue), 1300 (turquoise), and 2400 K (green), all at $\log(g) = 4.5$. Solid lines show results from the `PICASO 4.0` framework, while dashed lines correspond to the `Sonora Diamondback` grid. The condensation curves for each cloud species included in `Sonora Diamondback` are shown as well.

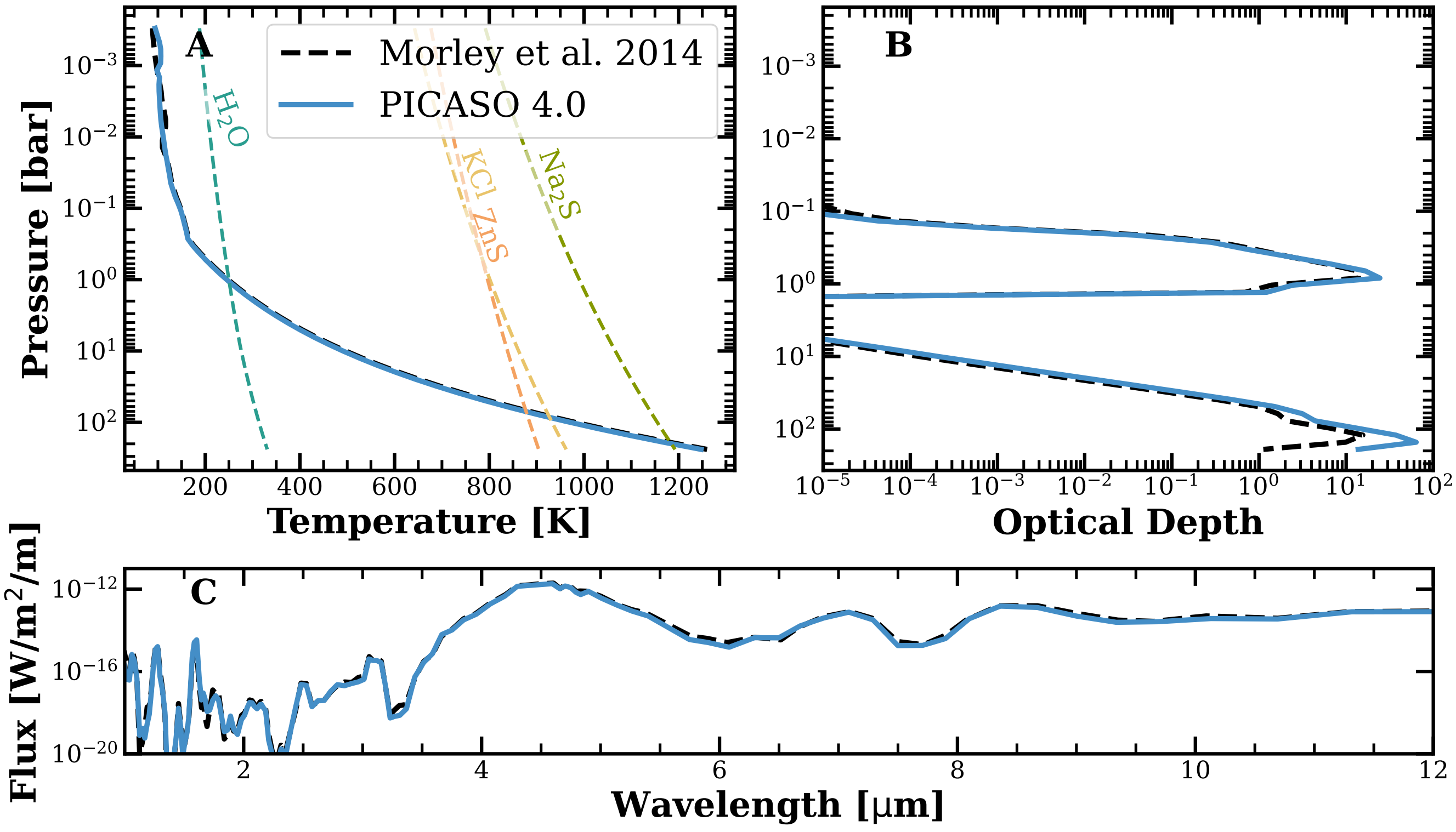

**Figure 9.** Comparison between a brown dwarf model generated using `PICASO` and the patchy-cloud model from C. V. Morley et al. (2014b), for an object with $T_{\mathrm{eff}} = 200$ K, $\log(g) = 4$, $f_{\mathrm{sed}} = 5$, and a cloud-hole fraction $f_{\mathrm{hole}} = 0.5$. Panel (A) shows the resulting pressure–temperature profiles from each model along with the respective cloud condensation curves. Panel (B) displays the cloud optical depth profiles at ∼4 μm. Panel (C) compares the thermal emission spectra.[18]

### 3.1. Brown Dwarf Cloud Profiles—Sonora Diamondback

Sonora Diamondback (C. V. Morley et al. 2024) presented a grid of models that included $MgSiO_3$, $Mg_2SiO_4$, Fe, and $Al_2O_3$ clouds for brown dwarfs with effective temperatures between 900 and 2400 K. These models were generated using `EGP` and A. S. Ackerman & M. S. Marley's (2001) cloud models. To demonstrate the new self-consistent cloud modeling feature in `PICASO 4.0`, we generated three different atmospheric profiles to compare with the Sonora Diamondback models and `Virga v2`. We generated cloudy models for 900, 1300, and 2400 K brown dwarfs with a surface gravity of log $(g) = 4.5$ and $f_{\mathrm{sed}} = 8$.

As shown in Figure 8, `PICASO` is able to reproduce the thermal structures of the cloudy Sonora Diamondback models. The small deviations we observe arise primarily from updates to the chemistry tables and from our different treatment of $K_{zz}$ in the radiative zone (see Section 2.4). In particular, the model at 2400 K exhibits the largest cloud-profile discrepancy, amounting to roughly 5% difference in optical depth compared to the original Sonora Diamondback profile. The observed differences in optical depth profiles are due to a slight variation in the $T(P)$ profile, where the PICASO profile is slightly cooler in the region where the clouds condense, changing where the cloud deck begins and the amount of the condensate available to condense, but overall it is a good match given the updates made in PICASO 4.0.

[18] Patchy Cloud Tutorial

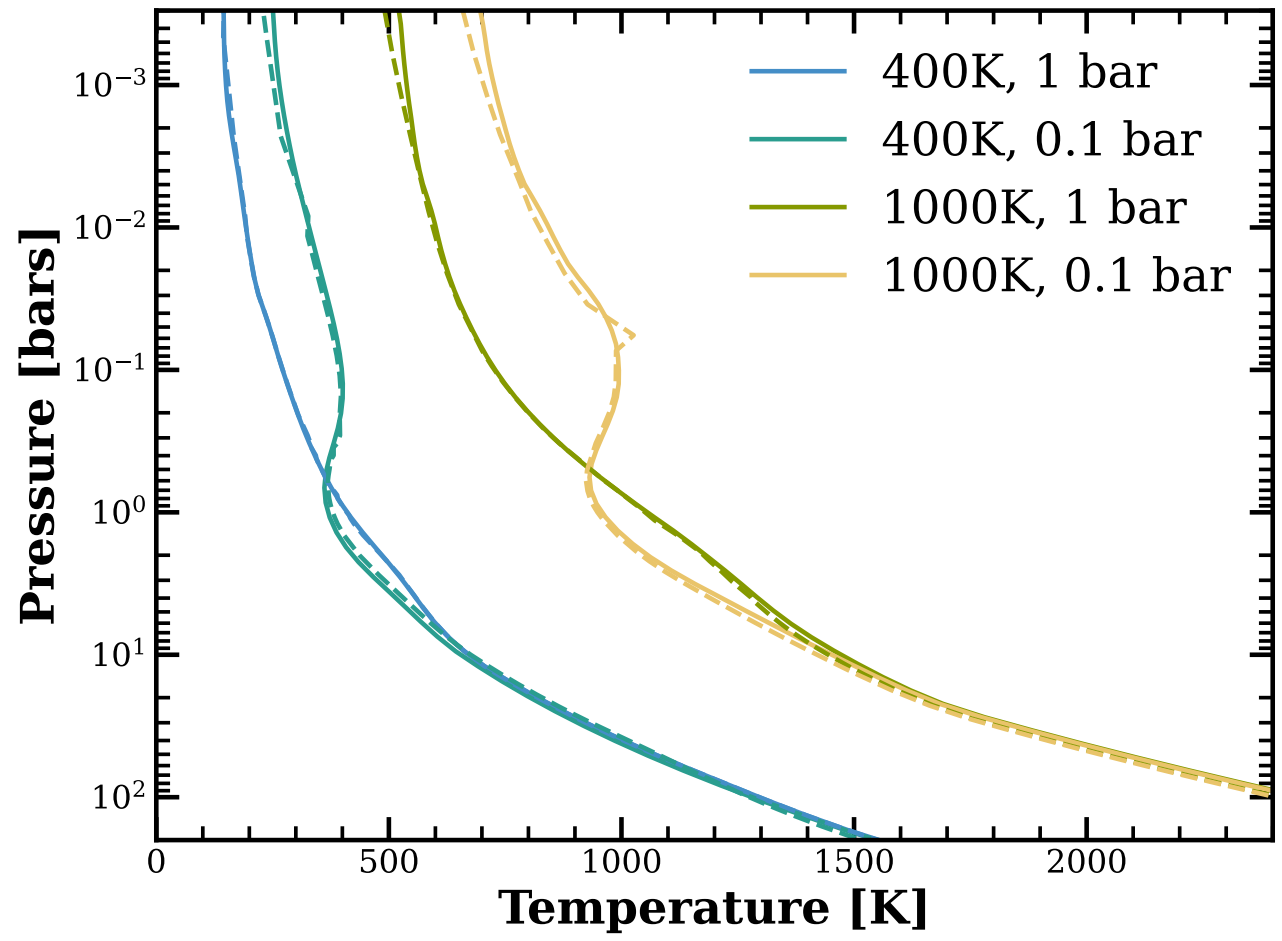

**Figure 10.** Temperature–pressure profiles generated with energy injected into the atmosphere for brown dwarfs with $T_{\rm eff}$ = 400 and 1000 K with the Chapman function centered at 0.1 and 1 bar. Solid lines show results from PICASO, while dashed lines represent the corresponding profiles from C. V. Morley et al. (2014a).[19]

### 3.2. Brown Dwarf Cloud Profiles—Patchy Clouds

To validate our patchy-cloud implementation, we benchmark our models against those from C. V. Morley et al. (2014b), which include clouds composed of $H_2O$, $Na_2S$, KCl, MnS, ZnS, and Cr. We generate self-consistent atmospheric profiles for a 200 K brown dwarf with a surface gravity of $\log g = 4$, assuming a sedimentation efficiency of $f_{\rm sed} = 5$ and a cloud-hole fraction of $f_{\rm hole} = 0.5$ (50% cloudy). As shown in Figure 9, our patchy-cloud treatment reproduces the thermal structure from C. V. Morley et al. (2014b) closely. The cloud optical depth for the deep salt and sulfur clouds differs by a factor of 4, while the $H_2O$ cloud optical depth profiles are nearly identical. This leads to an excellent match in the thermal emission spectra, where the largest difference between the spectra is only 0.28%.

### 3.3. Brown Dwarf Energy Injection

C. V. Morley et al. (2014a) generated atmospheric models with artificially injected energy to simulate hot-spot variability. We use their atmospheric profiles to benchmark the energy-injection functionality in PICASO. Using the energy_injection method, we compute pressure–temperature profiles for brown dwarfs with $T_{\rm eff}$ = 400 and 1000 K, $\log g = 4$, solar metallicity, and solar C/O ratios. Energy is deposited using a Chapman function as described in Section 2.6. Following C. V. Morley et al. (2014a), we inject a total of $7.26 \times 10^5$ erg cm$^{-2}$ s$^{-1}$ for the 400 K model and $2.84 \times 10^7$ erg cm$^{-2}$ s$^{-1}$ for the 1000 K model, distributed over one scale height centered at 0.1 and 1 bar. As shown in Figure 10, the resulting PICASO thermal structures almost perfectly match the original profiles. A small kink in the published 1000 K profile near 0.05 bar is absent in the PICASO result due to improved convergence. Overall, this comparison validates the ability to produce models in RCE with artificial energy injected into the atmosphere using PICASO.

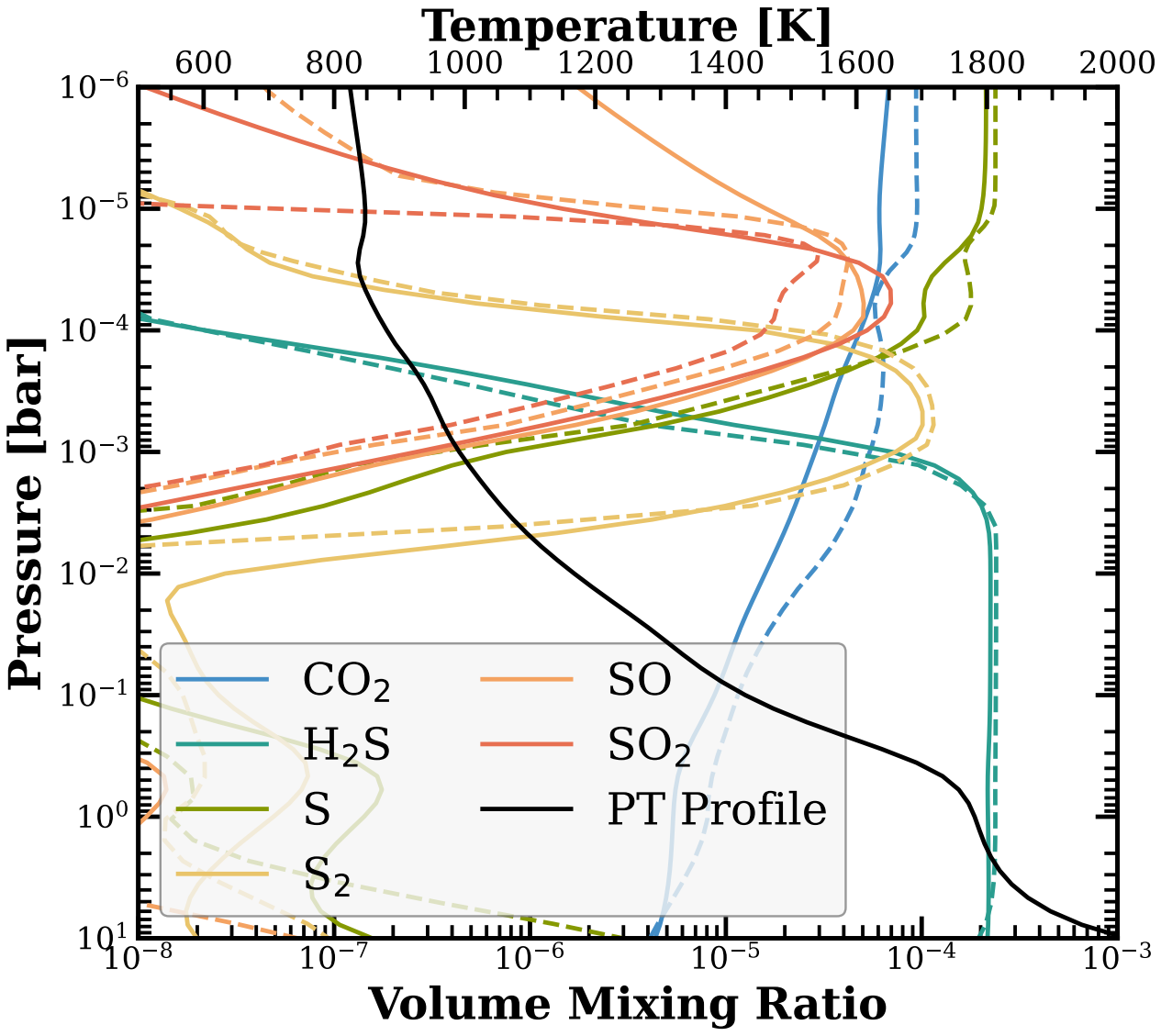

**Figure 11.** Chemical abundance profiles in the atmosphere of WASP-39b with photochemically produced species like $SO_2$ present in the upper atmosphere modeled by PICASO with photochem (solid) compared to the ATMO abundances at the evening terminator (dashed) from S.-M. Tsai et al. (2023). The PICASO thermal structure of WASP-39b is shown in black.[20]

### 3.4. WASP-39 b Self-consistent Photochemistry + Climate

To benchmark the coupled RCPE capabilities of PICASO 4.0, we begin with the published atmospheric results for WASP-39 b from the JWST Transiting Exoplanet Community ERS program (S.-M. Tsai et al. 2023), which identified photochemically produced $SO_2$ in the planet's upper atmosphere. We reproduce this scenario in PICASO by generating a self-consistent atmospheric profile with photochemistry (with inputs specified via the photochem_init_args input dictionary), specifying the pressure grid and stellar irradiation appropriate for WASP-39 b. Figure 11 shows the resulting abundances of key species, including $SO_2$, compared to results from ATMO published in S.-M. Tsai et al. (2023) for the evening terminator. The ATMO calculation (and others in S.-M. Tsai et al. 2023) specifically targets the morning and evening terminators, adopting a solar zenith angle of 83° and diurnal averaging factor of 1 for their photolysis rate calculations. Furthermore, they use $P$–$T$ profiles for each terminator from a three-dimensional climate model. Our coupled PICASO-photochem simulation instead attempts to simulate a one-dimensional global-average for efficient heat redistribution (60° solar zenith and a 0.5 diurnal averaging factor). These differences in model setup largely explain the marginal discrepancy for each abundance in Figure 11.

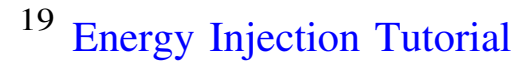

[19] Energy Injection Tutorial

[20] Photochemistry Tutorial

## 4. Future Developments

PICASO is completely open source, and we encourage the community to participate in developing the package. Here, we discuss upcoming updates to PICASO that are currently in development and will expand its capabilities, both in computational efficiency and scientific applicability.

### 4.1. PICASO 5.0: GPU-enabled Modeling

Forward modeling provides physically self-consistent atmospheric structures but can be computationally expensive, especially when producing grids of models. A clear, cloud-free brown dwarf atmosphere can typically be computed in roughly 10 minutes, whereas a cloudy model may require one to two hours, depending on the initial conditions set by the user (e.g., a very deep radiative convective boundary guess). This computational cost often becomes the limiting factor in efficiently interpreting observations. To address this, PICASO 5.0 will be GPU enabled. Leveraging GPUs will allow the most computationally intensive components, such as the radiative transfer, to be parallelized. Existing GPU-accelerated radiative transfer frameworks, such as HELIOS (M. Malik et al. 2017) and ExoJax (H. Kawahara et al. 2022; H. Yama et al. 2026), have already demonstrated the substantial performance gains achievable with this approach. Incorporating GPU support into PICASO will not only accelerate current workflows but will also create opportunities for future expansions, including the integration of more detailed physical or chemical processes and computationally demanding models, such as general circulation models and full line-by-line calculations.

### 4.2. PICASO 6.0: Terrestrial Planet Climates

Thanks to JWST, we can now characterize the atmospheres of many of these rocky exoplanets in far greater detail. The 500 hr JWST Rocky Worlds DDT program is now underway, alongside numerous other JWST efforts focused on terrestrial exoplanets that have surpassed over 1600 dedicated hours. These observations will directly inform future capabilities, especially those planned for the Habitable Worlds Observatory. PICASO is currently used to forward model the spectra of planets, including terrestrials. One minor PICASO 4.0 update we did not mention was the lifting of the emissivity $= 0$ constraint that had previously been hard-coded (C. M. Rooney et al. 2024). PICASO had previously included, as optional input, a wavelength-dependent or constant surface albedo. It also included flags for hard-surface or high-pressure boundary conditions. However, it did not inherently assume a surface emissivity $= 1 -$ albedo. This is a very minor change adopted in PICASO 4.0 and has already enabled the thermal emission spectra of some planets to be published (J. Adams Redai et al. 2025, e.g., GJ 375 b). Our current climate framework applies to substellar objects with H/He-dominated atmospheres; development is still needed to accurately model the secondary atmospheres expected on terrestrial planets, which are often carbon-, nitrogen-, or oxygen-dominated. PICASO 6.0 will incorporate updated adiabats, opacities, flexible chemistry, and heat-redistribution schemes tailored for rocky planets. Together, these advances will expand PICASO's climate modeling applicability beyond gas giants and substellar objects, enabling consistent modeling across a much broader range of planetary environments.

## 5. Conclusions

In this paper we have described the latest release of PICASO, which implements a number of major updates to the one-dimensional RCE climate modeling framework for H/He rich substellar objects.

1. We incorporated the new chemistry tables that have expanded the $P$, $T$ grid to 2121 points with additional abundances included for C, O, Mg, $Mg^+$, Si, $Fe^+$, Ti, $Ti^+$, and $C^+$.
2. The framework now supports multiple options for chemical disequilibrium, including rainout chemistry, cold trapping (also applicable in chemical equilibrium), and the selective removal of $PH_3$.
3. Our models can now account for photochemical effects through full integration with photochem.
4. By coupling to Virga, we can generate self-consistent cloudy atmospheres.
5. We include options for patchy clouds, allowing users to specify the fractional cloud coverage and explore heterogeneous cloud structures.
6. We implemented updated opacity datasets for $CH_4$, $NH_3$, and $SO_2$, alongside more adaptable methods for applying opacities within the climate solver.
7. We computed a new set of correlated-$k$ coefficients for each of the 2121 chemistry grid points and individual molecules.
8. Finally, we added a parameterized energy-injection capability that enables controlled heating of selected atmospheric layers, allowing users to explore a range of external and internal energy sources.

All these updates were motivated by the growing need to interpret atmospheres shaped by nonequilibrium chemistry, clouds, and additional heating-features that have been revealed in JWST datasets and will be essential to prepare for upcoming missions such as Roman and the ground-based ELTs.

We have demonstrated that the expanded PICASO framework can reproduce results from a variety of previously published models, including cloud-bearing brown dwarf atmospheres from Sonora Diamondback (C. V. Morley et al. 2024), patchy clouds from C. V. Morley et al. (2014b), energy-injected profiles from C. V. Morley et al. (2014a), and photochemical studies of WASP-39 b from S.-M. Tsai et al. (2023). These comparisons provide confidence that the new modules behave in a physically consistent manner with previous results.

PICASO remains an actively developed, open-source tool, and this release is intended to support increasingly diverse applications-from detailed case studies of individual objects to the construction of large model grids spanning wide parameter ranges. We have also released updated documentation and tutorials for each feature we presented in this paper, along with recommendations for best practices on GitHub. These tutorials are designed to detail all the required inputs for each functionality, the workflow, and how to interpret the outputs.

Upcoming updates to PICASO include GPU-enabled modeling to increase the computational efficiency for the climate models and expanding the applicability of the climate models beyond H/He-dominated atmospheres to terrestrial planets.

## Acknowledgments

The authors would like to thank the referee for their helpful comments, which improved the manuscript. We are grateful to the participants in our PICASO sessions at the annual meeting and hackathon, who helped test notebooks, identify minor bugs, and improve overall usability: Aditya Sengupta, Anna Gagnebin, Ben Lew, Benjamin Liberles, Kayla Smith, Laura Mayorga, Madeline Lessard, Melanie Rowland, Sarah Moran, Shelby Courreges, Tiffany Kataria, Yinan Zhao, and Zarah Brown. J.M. acknowledges support from the National Science Foundation Graduate Research Fellowship Program under grant No. DGE 2137420. S.M. is supported by the Heising-Simons Foundation through the 51 Pegasi b postdoctoral fellowship. N.E.B. acknowledges support from NASA's Interdisciplinary Consortia for Astrobiology Research (grant No. NNH19ZDA001N-ICAR) under award number 19-ICAR19_2-0041. This material is based on work supported by the National Aeronautics and Space Administration under grant No. 80NSSC24K0958 for the NASA XRP program. Support for programs JWST-AR-01977.004, JWST-GO-02507.004-A, and JWST-GO-2243 was provided by NASA through a grant from the Space Telescope Science Institute, which is operated by the Association of Universities for Research in Astronomy, Incorporated, under NASA contract NAS5-26555.

*Software:* PICASO (N. E. Batalha et al. 2019; S. Mukherjee et al. 2023; N. Batalha et al 2026), Virga (N. Batalha et al. 2020; N. E. Batalha et al. 2026; S. E. Moran et al. 2025), photochem (N. F. Wogan et al. 2025a), Jupyter (T. Kluyver et al. 2016), NumPy (S. v. D. Walt et al. 2011), SciPy (P. Virtanen et al. 2020), and Matplotlib (J. D. Hunter 2007).

## Author Contributions

J.M. implemented the major updates to PICASO 4.0 and led the manuscript and code release. N.E.B. compiled the new opacities and correlated-*k* tables, restructured the climate code, and managed the integration of updates. N.F.W. wrote the original photochem package and worked on the PICASO integration. S.M. wrote PICASO 3.0, helped with benchmarking and implementation of photochem into PICASO. C.V. did all the calculations for the new chemistry tables. K.L. C. computed new $NH_3$ opacities. C.V.M., M.S.M., and J.J.F. provided the guidance needed to interpret the original Fortran code, EGP, and benchmark tests. P.G. and I.M. contributed to critical bug fixes and testing. All authors contributed to writing the manuscript.

## ORCID iDs

James Mang https://orcid.org/0000-0001-5864-9599
Natasha E. Batalha https://orcid.org/0000-0003-1240-6844
Caroline V. Morley https://orcid.org/0000-0002-4404-0456
Nicholas F. Wogan https://orcid.org/0000-0002-0413-3308
Sagnick Mukherjee https://orcid.org/0000-0003-1622-1302
Channon Visscher https://orcid.org/0000-0001-6627-6067
Mark S. Marley https://orcid.org/0000-0002-5251-2943
Jonathan J. Fortney https://orcid.org/0000-0002-9843-4354
Katy L. Chubb https://orcid.org/0000-0002-4552-4559
Peter Gao https://orcid.org/0000-0002-8518-9601
Isaac Malsky https://orcid.org/0000-0003-0217-3880